# Coherence of quantum states after noiseless attenuation

S.U. Shringarpure, C.M. Nunn, T.B. Pittman, and J.D. Franson
*Physics Department, University of Maryland Baltimore County, Baltimore, Maryland 21250 USA*

Attenuating a quantum state using a beam splitter will introduce noise and decoherence. Here we show that heralding techniques can be used to attenuate Schrödinger cat states and squeezed vacuum states without any noise or decoherence [Mičuda et al., Phys. Rev. Lett. **109**, 180503 (2012)]. Noiseless attenuation also preserves quantum interference effects in nonclassical states such as squeezed vacuum states.

## I. INTRODUCTION

Photon loss in the transmission of continuous-variable quantum states can produce a large amount of decoherence, which limits the usefulness of continuous-variable quantum states in quantum communication systems. These effects can be reduced by noiselessly attenuating the signal prior to transmission, followed by noiseless amplification after transmission [1]. In this paper, we analyze the degree of coherence of several kinds of continuous-variable quantum states after they have been noiselessly attenuated. We show that ordinary attenuation by a beam splitter would introduce a large amount of decoherence, but that noiseless attenuation preserves the coherence of the quantum states and their ability to produce quantum interference effects.

Noiseless amplification techniques have been studied and experimentally verified, using linear or nonlinear optical elements combined with heralding techniques [2-5]. These probabilistic devices avoid the noise that is always introduced by deterministic, phase-preserving linear amplifiers [6, 7]. Similarly, the inverse transformation of noiseless attenuation can be implemented using several kinds of nondeterministic devices [1, 8]. The effect of noiseless attenuation can be described by the nonunitary operator $\nu^{\hat{n}}$, where the parameter $\nu$ can have values between 0 and 1. This transforms an input state of the form $\sum_n c_n |n\rangle$ into $\varepsilon \sum_n c_n \nu^n |n\rangle$, where $\varepsilon$ is a suitable normalization constant. In addition to reducing the average photon number [9], we will show that such a device is truly "noiseless" in the sense that it preserves the coherence of several nonclassical states of interest.

We will analyze the effects of a noiseless attenuator implemented using a beam splitter and conditional measurements (heralding) [1], as illustrated in Fig. 1. Noiseless attenuation can also be achieved using an optical parametric amplifier and heralding techniques [8]. Ordinary attenuation by a beam splitter (without heralding) will leave which-path information in the environment, which produces decoherence and a reduction in quantum interference. Heralding on zero photons in the upper output path of Fig. 1 eliminates any which-path information in the environment. Several recent experiments have demonstrated the feasibility of heralding on the detection of zero photons [10-16].

We will use the Wigner distributions [17] of the states as the primary tool for monitoring their evolution, since negative regions of the Wigner distribution are an indicator of nonclassicality and they can be used to test for any decoherence due to attenuation. In addition, the Wigner distribution is a useful tool since it can be reconstructed using homodyne measurements [18, 19].

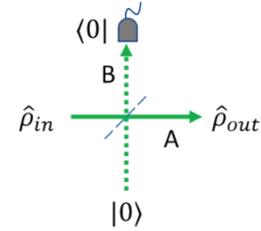

FIG. 1. A noiseless attenuator implemented using a beam splitter combined with heralding on the presence of zero photons in one of the output ports. The conditional measurement is represented by a projection $\langle 0|$ on the reflected mode.

The rest of the paper is as follows. In Sec. II, the nonclassicality of Schrödinger cat states is shown to be preserved under the action of a noiseless attenuator. In Sec. III, we consider the noiseless attenuation of single mode squeezed vacuum (SMSV) states. Section IV expands the analysis to two-mode states, including the effect of noiseless attenuation on the quantum interference of two SMSV states using a Mach-Zehnder interferometer. Section V deals with the effects of limited detector efficiency. A Summary and Conclusions are provided in Sec. VI.

## II. SCHRÖDINGER CAT STATES

Schrödinger cat states are a superposition of macroscopically distinguishable states. In the context of quantum optics, they are usually assumed to be a superposition of two coherent states [20]. For a coherent state, the Wigner distribution [17] is a Gaussian distribution centered at the corresponding amplitude. A cat state shows additional oscillations in between the Gaussians of the individual coherent states, as can be seen in Fig. 2. These oscillations are due to quantum interference between the two components of the cat state. The interference also gives rise to negative regions of the Wigner distribution, which is an indicator of the nonclassical nature of the state [21].

It is well known that photon loss from a Schrödinger cat state will leave which-path information in the environment, which suppresses the interference pattern in the Wigner distribution [22, 23]. If a noiseless attenuator is to be truly "noiseless," it must preserve the oscillations in the Wigner distribution. We will analyze the effects of the noiseless attenuator shown in Fig. 1, where the measurement of no photons in one of the output modes heralds the successful generation of the attenuated output in the other path.

We will assume that the input to the noiseless attenuator is an even cat state given by

$$|\psi_{cat}\rangle = \frac{|\alpha\rangle + |-\alpha\rangle}{\sqrt{2\left(1+e^{-2|\alpha|^2}\right)}}, \quad (1)$$

Here $\alpha$ is a real parameter and $|\alpha\rangle$ is a coherent state with that amplitude. A coherent state is given in the number basis by

$$|\alpha\rangle = e^{-|\alpha|^2/2} \sum_{n=0}^{\infty} \frac{\alpha^n}{\sqrt{n!}} |n\rangle. \quad (2)$$

If we consider relatively small amplitude cat states then, to a good approximation, we only need to keep a small number of photon number states $|n\rangle$. All of the subsequent numerical calculations were performed using an initial value of $\alpha = 2$ and keeping the first 20 photon number terms.

The wave function of the initial cat state in the coordinate representation [24] is thus a superposition of the wave functions $\psi_n(x)$ of the corresponding photon number states

$$\psi_{cat}(x) = \sum_n c_n \psi_n(x). \quad (3)$$

The coefficients $c_n$ can be obtained from Eqs. (1) and (2) as is described in more detail in the Appendix. The Wigner distribution for a pure state $|\psi\rangle$ in units where $\hbar = 1$ is then given [10] by the transformation

$$W(x,p) = \frac{1}{2\pi} \int_{-\infty}^{+\infty} dy\, e^{-ipy} \psi^*\!\left(x - \frac{y}{2}\right) \psi\!\left(x + \frac{y}{2}\right). \quad (4)$$

Figure 2 shows the Wigner distribution of the input cat state.

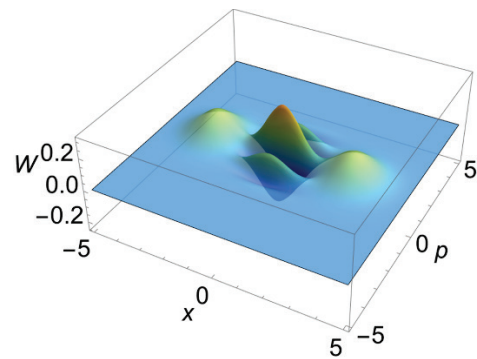

FIG. 2. Wigner distribution of the input cat state with $\alpha = 2$. The oscillations near the origin are due to quantum interference between the two coherent states in Eq. (1). The fact that the Wigner distribution has negative regions indicates that the state is nonclassical. (Dimensionless units.)

The beamsplitter transformation used to represent the input photon creation operators in terms of the output operators was chosen to be

$$B = \begin{pmatrix} t & ir \\ ir & t \end{pmatrix}, \quad (5)$$

where $t$ and $r$ are the transmissivity and the reflectivity of the beam splitter. Equation (5) is equivalent to several commonly used beam splitter transformations with the addition of different phases at its input and output modes.

### A. Ordinary attenuation

The beam splitter shown in Fig. 1 can couple photons into the output path labeled A as well as the auxiliary mode labeled B, which can be thought of as the environment. In the photon number basis, the state of the system after the beam splitter can be written as

$$|\psi_{out}\rangle = \sum_{n_a} \sum_{n_b} c(n_a, n_b) |n_a\rangle |n_b\rangle. \quad (6)$$

Here the coefficients $c(n_a, n_b)$ can be found using Eqs. (1), (2), and (5), while $|n_a\rangle$ and $|n_b\rangle$ correspond to states with $n_a$ and $n_b$ photons in the two output modes. The details of the calculations are described in the Appendix.

Since the number of photons in the environment is not measured in an ordinary attenuator, we need to take a partial trace over the environment. We will denote the projection onto the state with $n_b$ of photons in mode b as $|\psi_{out \mid n_b}\rangle$, which is given by

$$|\psi_{out \mid n_b}\rangle = \sum_{n_a} c(n_a, n_b) |n_a\rangle. \quad (7)$$

The density matrix of the mixed state after tracing over mode $b$ is given by

$$\hat{\rho}_{out} = \sum_{n_b} |\psi_{out \mid n_b}\rangle \langle \psi_{out \mid n_b}|. \quad (8)$$

The trace operation represents the decoherence due to loss of information into the environment.

The Wigner distribution of the mixed state after the partial trace is then given by

$$W_{out} = \sum_{n_b} W_{out \mid n_b}. \quad (9)$$

Here $W_{out \mid n_b}$ is the Wigner distribution of state $|\psi_{out \mid n_b}\rangle$. Note that $|\psi_{out \mid n_b}\rangle$ is an un-normalized state in our notation.

Figure 3 shows the Wigner distribution of the mixed state after tracing over the environment. The peaks corresponding to the original coherent states have been moved closer to the origin due to the overall attenuation. In addition, the oscillations near the origin have been reduced and are not as negative as before attenuation, which indicates a loss of decoherence and less nonclassical behavior.

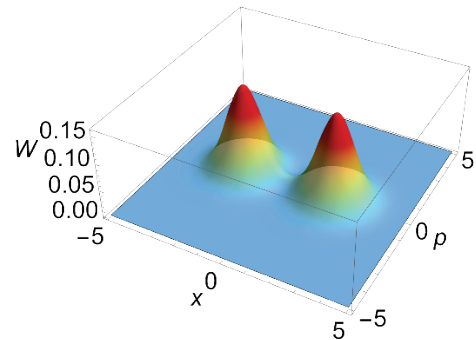

FIG. 3. Wigner distribution of the output state after a Schrodinger cat state has passed through an ordinary beam splitter. Here, the reflectivity of the beam splitter was arbitrarily chosen to be $r = \sqrt{0.5}$ (a 50-50 beam splitter). The oscillations near the origin have been reduced due to decoherence arising from the which-path information left in the environment. (Dimensionless units.)

### B. Noiseless attenuation

In the previous section, we considered the case in which there was no heralding based on the number of photons that were coupled into the auxiliary mode (environment), which reduces the amount of quantum interference. Now we will analyze the output of a noiseless attenuator in which the output is only accepted when no photons are found in the auxiliary mode.

With postselection of that kind, the Wigner distribution of the output mode is obtained by keeping only the $n_b = 0$ term in Eq. (9). After renormalization, this gives

$$W_{out}^{(heralded)} = \frac{W_{out\,|\,0_b}}{\langle \psi_{out\,|\,0_b} | \psi_{out\,|\,0_b} \rangle}, \qquad (10)$$

where

$$|\psi_{out\,|\,0_b}\rangle = \sum_n c_n t^n |n\rangle. \qquad (11)$$

The results are plotted in Fig. 4, where it can be seen that the oscillations in the Wigner distribution have been restored. The negativity of the Wigner distribution is also similar to that of the original state in Fig. 2. At the same time, the peaks due to the two coherent states have been moved closer to the origin as a result of the attenuation.

For comparison, the Wigner distribution of an exact Schrödinger cat state corresponding to a superposition of coherent states with a reduced amplitude of $\alpha = 1.414$ is plotted in Fig. 5. It can be seen that the Wigner distributions in Fig. 4 and 5 are the same, as can be shown analytically as well. The noiseless attenuation of a cat state is equivalent to simply reducing the amplitude of the coherent states in Eq. (1) while maintaining the coherence of their superposition. The amplitude of coherent states in the cat state at the output is given by $t\alpha$.

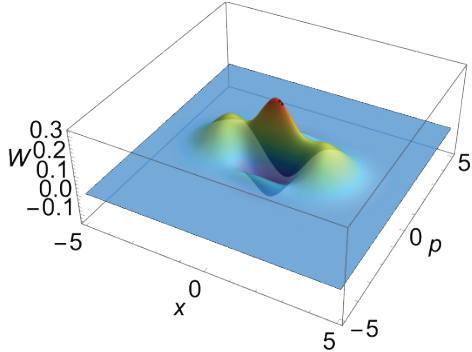

FIG. 4. Wigner distribution of the state of the output mode after noiseless attenuation of a Schrodinger cat state. Here the output was postselected for events in which no photons were observed in the auxiliary mode. The oscillations near the origin are much larger than is the case for ordinary attenuation in Fig. 3, which shows that the coherence of the state has been maintained. (Dimensionless units.)

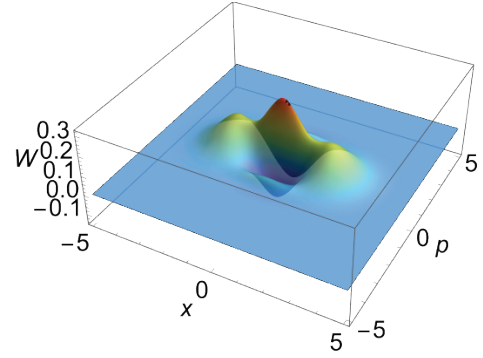

FIG. 5. Wigner distribution of an exact Schrödinger cat state with a coherent-state amplitude of $\alpha = 1.414$. Comparing these results with those of Fig. 4 shows that noiseless attenuation of a cat state is equivalent to simply reducing the amplitude of the two coherent states in Eq. (1). (Dimensionless units.)

The success probability $P_{success}$ is state dependent and given by

$$P_{success} = \langle \psi_{out\,|\,0_b} | \psi_{out\,|\,0_b} \rangle = \sum_n |c_n|^2 t^{2n}. \qquad (12)$$

If the input state can be approximated by an expansion in the Fock basis with a maximum of $N$ photons, then it is evident from Eq. (12) that $P_{success} \geq t^{2N}$; i.e., it is lower bounded [1]. This is a good approximation for weak cat states and squeezed vacuum states and for some $N$. For the example shown in Fig. 4, $P_{success} = 0.14$.

## III. SINGLE-MODE SQUEEZED VACUUM

We saw in the previous section that a noiseless attenuator preserves the coherence of a Schrödinger cat state. We will now consider another example in which a single mode squeezed vacuum (SMSV) state with squeezing along the x quadrature is passed through a noiseless attenuator. The initial SMSV state is given [25] by

$$|\psi_{SMSV}\rangle = \frac{1}{\sqrt{\cosh \xi}} \sum_{n=0}^{\infty} \sqrt{\binom{2n}{n}} \left(-\frac{\tanh \xi}{2}\right)^n |2n\rangle, \qquad (13)$$

where $\xi$ is a parameter related to the strength of the interaction in a $\chi^{(2)}$ medium. States of this kind can be produced using parametric down-conversion [26] and they are widely used in many applications.

The Wigner distribution of a single-mode squeezed vacuum state is described by a Gaussian of the form [17]

$$W(x, p) = A \exp\left[-\left(sx^2 + \frac{p^2}{s}\right)\bigg/2\sigma^2\right], \qquad (14)$$

as illustrated in Fig. 6. The squeezing parameter $s$ is related to $\xi$ by $\xi = \ln\sqrt{s}$, while the width $\sigma$ has the same value as in an ordinary vacuum state where $s = 1$ and $\sigma = 1/\sqrt{2}$. The uncertainty in one direction of phase space is reduced at the expense of an increased uncertainty in the orthogonal direction, as required by the uncertainty principle.

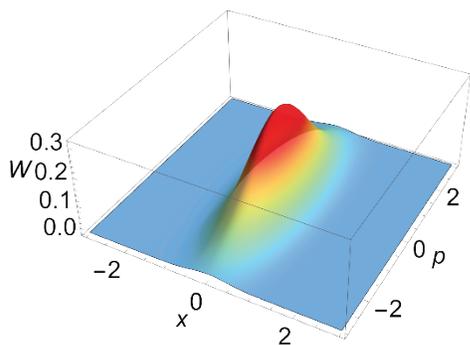

FIG. 6. Wigner distribution of a single mode squeezed vacuum state with a squeezing parameter of $s = 3$ (arbitrary units). Fitting the distribution with Eq. (14) gives $\sigma = 0.707$. (Dimensionless units.)

Figure 7 shows the Wigner distribution of the output state after a single mode squeezed vacuum state has passed through an ordinary attenuator consisting of a 50-50 beam splitter. In comparison, Fig. 8 shows the Wigner distribution after the state has passed through a noiseless attenuator with postselection on the auxiliary mode as discussed earlier. A reduction in the squeezing can be clearly seen in both cases. A fit to Eq. (12) gives $\sigma = 0.759$ and $s = 1.732$ for ordinary attenuation, while noiseless attenuation gives $\sigma = 0.707$ and $s = 1.667$. It can be seen that noiseless attenuation gives a state with lower uncertainty (noise) than is obtained using ordinary attenuation, although the difference is not as apparent as it is for a Schrödinger cat state. The new squeezing parameter in the state at the output is given by $\left[s(1+t^2)+(1-t^2)\right]\big/\left[s(1-t^2)+(1+t^2)\right]$. For the example shown in Fig. 7, $P_{success} = 0.90$.

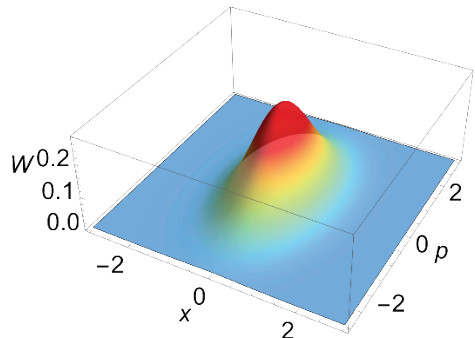

FIG. 7. Wigner distribution of a single mode vacuum state after it has passed through an ordinary attenuator consisting of a beam splitter with an arbitrarily chosen reflectivity $r = \sqrt{0.5}$. Fitting the distribution with Eq. (14) gives $\sigma = 0.759$ and $s = 1.732$. (Dimensionless units.)

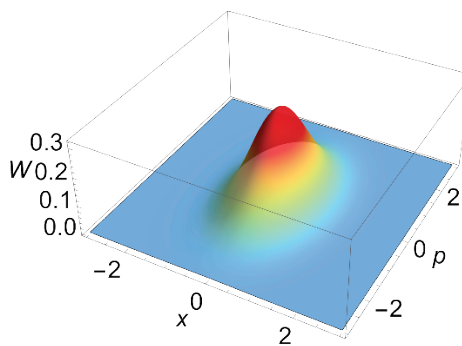

FIG. 8. Wigner distribution of a single mode squeezed vacuum state after noiseless attenuation using a beam splitter and heralding, as illustrated in Fig. 1. Fitting the distribution with Eq. (14) gives $\sigma = 0.707$, which is same as that for the original single-mode squeezed state in Fig. 6, along with a value of $s = 1.667$. (Dimensionless units.)

## IV. QUANTUM INTERFERENCE

Figure 4 shows that noiseless attenuation maintains the quantum interference that is responsible for the oscillations near the origin of the Wigner distribution of a Schrödinger cat state. In this section, we will use a Mach-Zehnder interferometer to give a more explicit demonstration of the effects of a noiseless amplifier. In an ordinary attenuator, which-path information left in the environment will produce a large decrease in the visibility of the interference pattern. A noiseless attenuator eliminates the which-path information and would be expected to maintain the coherence of the quantum interference.

The interferometer measurements of interest are illustrated in Fig. 9. A two-mode squeezed state is incident in the two input ports of the 50-50 beam splitter which is assumed to have the same form as in Eq. (5). It can be shown that this transformation generates two independent single-mode squeezed states in the two output modes [27]. These single-mode-squeezed states then pass through the noiseless attenuators placed in both paths, which consist of a beam splitter and heralding on zero photons in one of the output paths as in Fig. 1. Finally, the two beams are mixed on a second beam splitter to form a Mach-Zehnder interferometer. Coincidence measurements are performed on the two outputs.

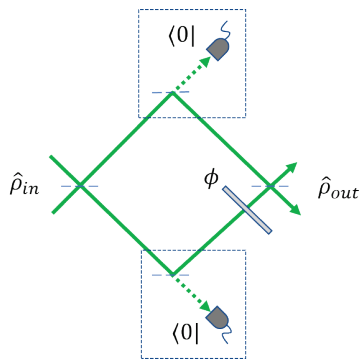

FIG. 9. A modified Mach-Zehnder interferometer that could be used to measure the amount of quantum interference between two states after noiseless attenuation. The input state is a two-mode squeezed state, which is transformed into two independent single mode squeezed states after passing through the first beam splitter on the left. The noiseless attenuators are shown enclosed in blue dashed boxes. A phase shift $\phi$ is applied in one path, after which the two modes are recombined at a second beam splitter on the right. The effects of quantum interference can be observed in coincidence measurements between the two output ports.

The incident two-mode squeezed state can be written in the number state basis in the form [28]

$$|\psi_{TMSV}\rangle = \frac{1}{\cosh\xi}\sum_{n=0}^{\infty}(-\tanh\xi)^{n}|n\rangle|n\rangle. \qquad (15)$$

Here $\xi$ is once again related to the strength of the squeezing interaction, which we arbitrarily assumed to have the value $\xi = 0.5$. For relatively small squeezing, it is sufficient to retain only the first few terms in a number-state expansion. Equation (15) can be used to calculate the effects of the first beam splitters, and the coincidence rate was calculated numerically using the same techniques as before.

With 100% reflective beam splitters (mirrors) in the two paths through the interferometer, Fig. 9 reduces to a standard Mach-Zehnder interferometer, and the calculated results show a visibility of 100% in the interference pattern. By reducing the reflectivity of the intermediate beam splitters, ordinary attenuation is introduced in both paths, adding noise. An arbitrarily chosen reflectivity of $r = \sqrt{0.5}$, for both beam splitters, gives a reduced visibility of ~90%. Noiseless attenuation is achieved by heralding on zero photons in the auxiliary modes, as shown in the dashed boxes of Fig. 9. This restores the interference visibility to 100%, as illustrated in Figure 10. For the example shown in Fig. 10, $P_{success} = 0.83$.

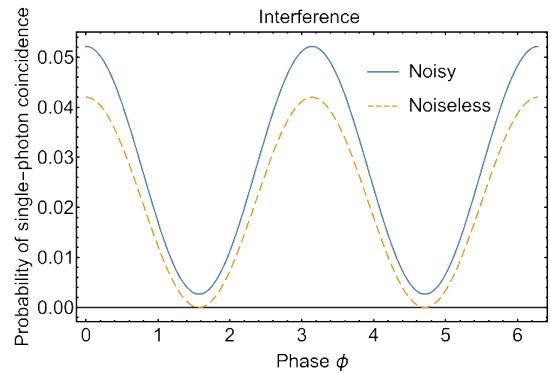

FIG. 10. Probability of a single-photon coincidence in the two output paths of the Mach-Zehnder interferometer of Fig. 9. The solid (blue) line shows the results for the case of ordinary attenuation, where no heralding on the auxiliary mode was performed. This reduces the visibility of the interference pattern to ~90%. The dashed (orange) line shows the results for noiseless attenuators, where the output was heralded on the presence of zero photons in the auxiliary mode; this gives a visibility of 100%. Both cases correspond values of $\xi = 0.5$ and $r = \sqrt{0.5}$, chosen arbitrarily. (Dimensionless units.)

These results show that noiseless attenuation does maintain the coherence required for quantum interference effects.

## V. DETECTOR EFFICIENCY

Up to this point, the single-photon detectors used in the heralding process were assumed to have 100% detection efficiency. Limited detection efficiency can have a significant effect on the output of the heralded detection process shown in Fig. 1, which will no longer be completely "noiseless" [16]. In this section, we will model the effects of limited detection efficiency using a perfect detector preceded by a beam splitter to simulate the effects of loss or detection inefficiency.

Figure 11 shows the effects of a noiseless detector on a Schrödinger cat state for several different values of the detector efficiency. It can be seen that the coherence of the cat state is completely maintained for a perfect detector, but that the performance of the device gradually degrades to that of an ordinary attenuator for a detection efficiency of 0. Intermediate values of the detection efficiency produce a reduction in the oscillations near the origin of the Wigner distribution, indicating a gradual decrease in the coherence of the output state. It can be seen from Fig. 11 (c) that a detection efficiency as low as 50% can still give significantly better performance than an ordinary attenuator. For an initial cat state, the Wigner function of the output state can be solved analytically for arbitrary detector efficiency (see the Appendix).

The dependence of the Mach-Zehnder interferometer of Fig. 9 on the detection efficiency is shown in Fig. 12, which is a plot of the visibility as a function of detector efficiency. There is a decrease in the visibility for lower efficiency photodetectors, which can be understood from the fact that a detector with limited efficiency does not completely rule out the possibility of which-path information being left in the environment, as in ordinary attenuation.

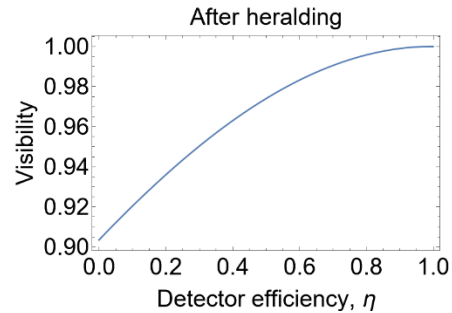

FIG. 12. Visibility of the quantum interference from the Mach-Zehnder interferometer of Fig. 9 as a function of the detector efficiency used in the heralding process. No which-path information is left in the environment and the visibility is 100% for a perfect detector. The visibility decreases for limited detection efficiency since the possibility of which-path information is not completely eliminated in that case. (Dimensionless units.)

The overall process is still noisy, and we are simply heralding on a suitable subset of the output states, which eliminates the terms that would have contributed to the noise. The ability to eliminate these outcomes, however, depends on the detector efficiency, which in turn has an effect on the amount of which-path information lost to the environment.

## VI. SUMMARY AND CONCLUSIONS

Ordinary attenuation of an optical quantum state using a beam splitter can produce decoherence due to which-path information left in the environment. Noiseless attenuation can be achieved using a beam splitter combined with heralding on those events in which no photons are present in the auxiliary mode (the environment), which eliminates the which-path information. It is interesting that this process can reduce the intensity of an optical signal without extracting any power from the system [1, 8].

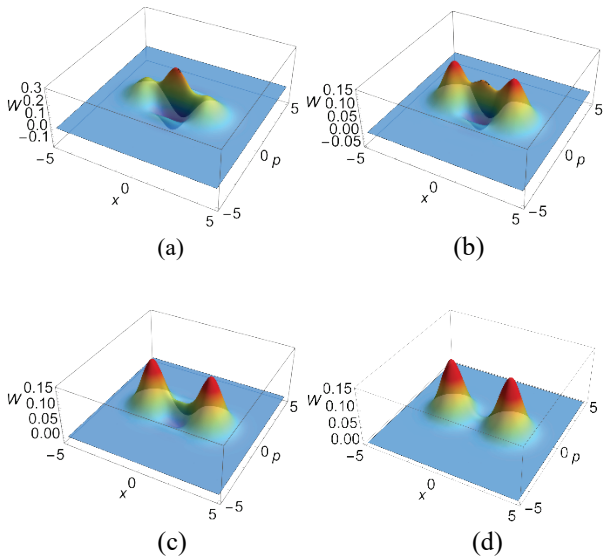

FIG. 11. Effect of using inefficient detectors for heralding on no photons for noiseless attenuation of cat states. Wigner distribution of the outputs when the efficiency is: (a) 100%, (b) 75%, (c) 50%, and (d) 0%. (Dimensionless units.)

In this paper, we showed that noiseless attenuators are truly "noiseless" in the sense that they do not reduce the coherence of the input state. We first considered the case of a Schrödinger cat state that has passed through a noiseless attenuator. The Wigner distribution of a cat state has characteristic oscillations near the origin that arise from the interference of its two constituent coherent states. The Wigner distribution also has negative regions, which demonstrates that the states are nonclassical. We showed that noiseless attenuation maintains both of these properties. We also found similar results for the case of a single-mode squeezed vacuum state, where noiseless attenuation maintained the width of the Gaussian Wigner distribution.

Quantum interference effects were investigated more directly by considering a Mach-Zehnder interferometer with noiseless attenuators in each arm and a two mode-squeezed vacuum state for the input. The visibility in the interference pattern from coincidence measurements was found to be maintained by noiseless attenuation, while it was substantially reduced by ordinary attenuation. Once again, this is due to the fact that the heralding process eliminates any which-path information left in the environment.

The effects of limited detection efficiency were also investigated. As might be expected, heralding using a detector with limited detection efficiency limits the ability of the heralding process to eliminate noise by eliminating those states that would leave which-path information in the environment.

These results may be of practical use in quantum communications systems based on continuous variables, where photon loss will result in the decoherence of nonclassical states. These effects can be reduced by noiselessly attenuating the signal before transmission, followed by noiseless amplification after transmission [1]. Our results show that noiseless attenuation can maintain the coherence of nonclassical states, but that detector efficiency will be an important consideration.

## ACKNOWLEDGEMENT

This work was supported in part by the National Science Foundation under Grant No. PHY-1802472 and PHY-2013464.

## APPENDIX

Some of the details of the calculations outlined in the text are presented in this Appendix.

We first consider the form of an even Schrödinger cat state. Combining Eqs. (2) and (3) gives

$$|\psi_{cat}\rangle = \frac{e^{-|\alpha|^2/2}}{\sqrt{2(1+e^{-2|\alpha|^2})}} \sum_{n=0}^{\infty} \frac{\alpha^n + (-\alpha)^n}{\sqrt{n!}} |n\rangle. \quad (A1)$$

Since the even terms are the only ones that contribute, we can introduce a new variable $k = n/2$, which gives

$$|\psi_{cat}\rangle = \frac{1}{\sqrt{\cosh|\alpha|^2}} \sum_{k=0}^{\infty} \frac{\alpha^{2k}}{\sqrt{(2k)!}} |2k\rangle. \quad (A2)$$

This expression gives the values of the coefficients $c_n$ that appear in Eq. (3). Rewriting the number states in terms of photon creation operators acting on vacuum gives

$$|\psi_{cat}\rangle = \frac{1}{\sqrt{\cosh|\alpha|^2}} \sum_{k=0}^{\infty} \frac{\alpha^{2k}}{(2k)!} \left(\hat{a}^\dagger\right)^{2k} |0\rangle. \quad (A3)$$

The cat state of Eq. (A2) passes through a beam splitter as illustrated in Fig. 1. We use the beam splitter transformation of Eq. (5) to relate the photon creation operators in the input mode A to those in the output modes A and B, which gives

$$\hat{a}^\dagger_{in} \to t\hat{a}^\dagger_{out} + ir\hat{b}^\dagger_{out}. \quad (A4)$$

We have chosen to represent the photon creation operators in modes A and B by $\hat{a}^\dagger$ and $\hat{b}^\dagger$, respectively. Inserting Eq. (A4) into Eq. (A3) gives the output state after the beam splitter:

$$|\psi_{cat}\rangle = \frac{1}{\sqrt{\cosh|\alpha|^2}} \sum_{k=0}^{\infty} \frac{\alpha^{2k}}{(2k)!} \left(t\hat{a}^\dagger + ir\hat{b}^\dagger\right)^{2k} |0\rangle|0\rangle. \quad (A5)$$

Using the binomial expansion and then applying the photon creation operators to the vacuum state, this can be rewritten as

$$|\psi_{cat}\rangle = \frac{1}{\sqrt{\cosh|\alpha|^2}} \sum_{k=0}^{\infty} \sum_{l=0}^{2k} \frac{\alpha^{2k}}{\sqrt{(2k-l)!l!}} \quad (A6)$$
$$\times (t)^{2k-l} (ir)^l |2k-l\rangle |l\rangle.$$

Introducing two new variables defined by $n_b = l$ and $n_a = 2k - l$ gives

$$|\psi_{cat}\rangle = \frac{1}{\sqrt{\cosh|\alpha|^2}} \sum_{n_a=0}^{\infty} \sum_{n_b=0}^{\infty} f(n_a, n_b) \quad (A7)$$
$$\times \frac{(t\alpha)^{n_a} (ir\alpha)^{n_b}}{\sqrt{n_a! n_b!}} |n_a\rangle |n_b\rangle,$$

where $f(n_a, n_b) = (n_a + n_b + 1) \mod 2$. This gives the final state of Eq.(6) where the coefficients given are

$$c(n_a, n_b) = \left[(n_a + n_b + 1) \mod 2\right] \frac{(t\alpha)^{n_a} (ir\alpha)^{n_b}}{\sqrt{n_a! n_b! \cosh|\alpha|^2}}. \quad (A8)$$

By setting $n_b = 0$ in Eq. (A8) and comparing with the coefficients in Eq.(A2) we see that heralding on zero photons in mode b would simply give another cat state with coherent states of amplitude $\alpha_{out} = t\alpha$ in mode a. This can also be noted by the transformation of the input state coefficients, $c_n$, to the output ones, $c_n t^n$. Therefore, the Wigner function of the output is

$$W_{out}^{(heralded)} = $$
$$\left[ e^{-(x+\sqrt{2}t\alpha)^2 - p^2} + e^{-(x-\sqrt{2}t\alpha)^2 - p^2} \right. \quad (A9)$$
$$\left. + 2e^{-(x^2+p^2)} \cos(2\sqrt{2}t\alpha p) \right] / 2\pi \left(1 + e^{-|t\alpha|^2}\right).$$

Using the same procedure for an input state consisting of the single mode squeezed vacuum of Eq. (11) instead of an even Schrödinger cat state gives a set of coefficients of the form

$$c(n_a, n_b)$$
$$= \left[(n_a + n_b + 1) \mod 2\right] \frac{(n_a + n_b)!}{\left(\frac{n_a + n_b}{2}\right)!}$$
$$\times \frac{\left(t\sqrt{\frac{-\tanh\xi}{2}}\right)^{n_a} \left(ir\sqrt{\frac{-\tanh\xi}{2}}\right)^{n_b}}{\sqrt{n_a! n_b! \cosh\xi}}. \quad (A10)$$
$$= \left[(n_a + n_b + 1) \mod 2\right] \left[\prod_{k=1}^{(n_a+n_b)/2} (2k-1)\right]$$
$$\times \frac{\left(t\sqrt{-\tanh\xi}\right)^{n_a} \left(ir\sqrt{-\tanh\xi}\right)^{n_b}}{\sqrt{n_a! n_b! \cosh\xi}}.$$

By setting $n_b = 0$ in Eq. (A10) and comparing with the coefficients in Eq.(13) we see that heralding on zero photons in mode b would simply give another squeezed vacuum state with the squeezing parameters of the output given by $\tanh\xi_{out} = t^2 \tanh\xi$, or alternatively $s_{out} = \left[s(1+t^2) + (1-t^2)\right] / \left[s(1-t^2) + (1+t^2)\right]$. This can also be noted by the transformation of the input state coefficients, $c_n$, to the output ones, $c_n t^n$. For a given initial $s$, $s_{out}(t)$ is a monotonically decreasing function of the transmittivity $t$. Therefore, the Wigner function of the attenuated output squeezed vacuum state is

$$W_{out}^{(heralded)} = \frac{1}{\pi} \times$$
$$Exp\left[ -\frac{s(1+t^2) + (1-t^2)}{s(1-t^2) + (1+t^2)} x^2 \right. \quad (A11)$$
$$\left. -\frac{s(1-t^2) + (1+t^2)}{s(1+t^2) + (1-t^2)} p^2 \right].$$

Equations (A9) and (A11) give analytic results for the case of a Schrödinger cat or a single mode squeezed state incident on a beam splitter. Although a similar calculation could be done for the case of a two-mode squeezed vacuum incident on a Mach-Zehnder interferometer as in Fig. 9, the analysis is more tedious and numerical solutions were used instead.

If heralding on zero is done using a detector of efficiency $\eta$, instead of setting $n_b = 0$ we need to use the density operator formalism and use the projector,

$$\hat{\Pi}_0 = \sum_{n_b=0}^{\infty} (1-\eta)^{n_b} |n_b\rangle\langle n_b|. \quad (A12)$$

This gives the un-normalized output state

$$\tilde{\rho}_{out}^{(heralded)} = Tr\left[\hat{\rho}_{in}\hat{\Pi}_0\right] = \sum_{n_b=0}^{\infty} (1-\eta^{n_b})|\psi_{out|n_b}\rangle\langle\psi_{out|n_b}|, \quad (A13)$$

If the input is an even cat state, then for even $n_b (\equiv 2k)$,

$$|\psi_{out|2k}\rangle$$
$$= \frac{(ir\alpha)^{2k}}{\sqrt{(2k)!}} \sqrt{\frac{\cosh(|t\alpha|^2)}{\cosh(|\alpha|^2)}} \left(\frac{1}{\sqrt{\cosh|t\alpha|^2}} \sum_{m=0}^{\infty} \frac{(t\alpha)^{2m}}{\sqrt{(2m)!}}|2m\rangle\right)$$
$$= \frac{(ir\alpha)^{2k}}{\sqrt{(2k)!}} \sqrt{\frac{\cosh(|t\alpha|^2)}{\cosh(|\alpha|^2)}} |\text{Even Cat}(t\alpha)\rangle, \quad (A14)$$

and similarly for odd $n_b (\equiv 2k+1)$,

$$|\psi_{out|2k+1}\rangle = \frac{(ir\alpha)^{2k+1}}{\sqrt{(2k+1)!}} \sqrt{\frac{\sinh(|t\alpha|^2)}{\cosh(|\alpha|^2)}} |\text{Odd Cat}(t\alpha)\rangle. \quad (A15)$$

Using Eqs. (A14) and (A15) in (A13) give the un-normalized state,

$$\tilde{\rho}_{out}^{(heralded)} = $$
$$\left[\sum_{k=0}^{\infty} \frac{(1-\eta)^{2k}(|ir\alpha|^2)^{2k}}{(2k)!}\right] \left[\frac{\cosh(|t\alpha|^2)}{\cosh(|\alpha|^2)}\right] \hat{\rho}_+ \quad (A16)$$
$$+ \left[\frac{\sinh(|t\alpha|^2)}{\cosh(|\alpha|^2)}\right] \hat{\rho}_-,$$

where $\hat{\rho}_+$ and $\hat{\rho}_-$ are the normalized density operators for even and odd cat states with amplitude $t\alpha$ respectively. This simplifies to the normalized output

$$\hat{\rho}_{out}^{(heralded)} = \frac{p_+\hat{\rho}_+ + p_-\hat{\rho}_-}{p_+ + p_-}, \quad (A17)$$

where

$$p_+ = \frac{\cosh(|t\alpha|^2)\cosh\left[(1-\eta)|ir\alpha|^2\right]}{\cosh(|\alpha|^2)}, \quad (A18)$$

and

$$p_- = \frac{\sinh(|t\alpha|^2)\sinh\left[(1-\eta)|ir\alpha|^2\right]}{\cosh(|\alpha|^2)}. \quad (A19)$$

The corresponding Wigner function is

$$W_{out}^{(heralded)} = \frac{p_+ W_+ + p_- W_-}{p_+ + p_-}, \quad (A20)$$

where

$$W_\pm = \left[e^{-(x+\sqrt{2}t\alpha)^2 - p^2} + e^{-(x-\sqrt{2}t\alpha)^2 - p^2}\right.$$
$$\left.\pm 2e^{-(x^2+p^2)}\cos(2\sqrt{2}t\alpha p)\right] / 2\pi\left(1 \pm e^{-|t\alpha|^2}\right). \quad (A21)$$

The success probability of the heralding is $p_+ + p_-$. It may be interesting to note that in this case, the noise remaining in the system is simply an odd cat state. In general, for other input states, the output heralded with an inefficient detector will have a more complicated noise term.